\pdfoutput=1
\documentclass[aps,prb,twocolumn,showpacs]{revtex4-1}
\usepackage{graphicx}
\usepackage{natbib}
\usepackage[utf8]{inputenc}
\usepackage{epstopdf}
\setcitestyle{super}
\usepackage{color}
\usepackage{amsmath}
\usepackage{tabularx}

\newcommand*{\citen}[1]{%
  \begingroup
    \romannumeral-`\x 
    \setcitestyle{numbers}%
    \cite{#1}%
  \endgroup
}

\usepackage{pdfpages}

\makeatletter
\AtBeginDocument{\let\LS@rot\@undefined}
\makeatother

\begin{document}

\preprint{APS/123-QED}
\date{\today}

\title{Evidence for electronically-driven ferroelectricity in a strongly correlated dimerized BEDT-TTF molecular conductor}

\author{Elena Gati$^1$}
\author{Jonas K.H. Fischer$^2$}
\author{Peter Lunkenheimer$^2$}
\author{David Zielke$^1$}
\author{Sebastian K\"ohler$^1$}
\author{Felizitas Kolb$^2$}
\author{Hans-Albrecht Krug von Nidda$^2$}
\author{Stephen M. Winter$^3$}
\author{Harald Schubert$^1$}
\author{John A. Schlueter$^{4,5}$}
\author{Harald O. Jeschke$^{3,6}$}
\author{Roser Valent\'{i}$^3$}
\author{Michael Lang$^1$}

\address{$^1$ Institute of Physics, Goethe-Universit\"{a}t, Max-von-Laue-Straße 1, 60438 Frankfurt am Main, Germany}
\address{$^2$ Experimental Physics V, Center for Electronic Correlations and Magnetism, University of Augsburg, 86159 Augsburg, Germany}
\address{$^3$ Institute for Theoretical Physics, Goethe-Universit\"{a}t, Max-von-Laue-Straße 1, 60438 Frankfurt am Main, Germany}
\address{$^4$ Division of Materials Research, National Science Foundation, Arlington, VA 22230, USA}
\address{$^5$ Materials Science Division, Argonne National Laboratory, Argonne, IL 60439, USA}
\address{$^6$ Research Institute for Interdisciplinary Science, Okayama University, 3-1-1 Tsushima-naka, Kita-ku, Okayama 700-8530, Japan}

\begin{abstract}
    By applying measurements of the dielectric constants and relative length changes to the dimerized molecular conductor $\kappa$-(BEDT-TTF)$_2$Hg(SCN)$_2$Cl, we provide evidence for order-disorder type electronic ferroelectricity which is driven by charge order within the (BEDT-TTF)$_2$ dimers and stabilized by a coupling to the anions. According to our density functional theory calculations, this material is characterized by a moderate strength of dimerization. This system thus bridges the gap between strongly dimerized materials, often approximated as dimer-Mott systems at 1/2 filling, and non- or weakly dimerized systems at 1/4 filling exhibiting charge order. Our results indicate that intra-dimer charge degrees of freedom are of particular importance in correlated $\kappa$-(BEDT-TTF)$_2$X salts and can create novel states, such as electronically-driven multiferroicity or charge-order-induced quasi-1D spin liquids.
\end{abstract}


\maketitle

\textit{Introduction. ---}
 Electronic ferroelectricity, where electrons play the role of the ions in conventional displacive ferroelectrics, has recently become an active area of research\cite{Ikeda05,vandenBrink08,Yamamoto10,Naka10,Ishihara10}. Characteristic of this novel type of ferroelectricity is that the polar state is controlled by electronic degrees of freedom of charge, spin and orbital nature, implying the intriguing possibility of cross-correlations with the material's magnetic properties.

 A key phenomenon for electronic ferroelectricity is charge order (CO), resulting from strong electronic correlations, and being ubiquitous in doped transition-metal oxides, such as high-$T_c$ cuprates \cite{Tranquada95} or manganites \cite{Mori98b}. Particularly clear examples of CO have been found in the families of TMTTF \cite{Chow00} and \mbox{BEDT-TTF} \cite{Mori98,Kuroki09,Kagawa13,Sasaki17} (in short ET) molecular conductors with 1/4-filled hole bands. It has been established that in these systems, CO and accompanying ferroelectric properties \cite{Monceau01,Nad06,Lunkenheimer15} result from the combined action of a strong onsite Coulomb repulsion $U$ along with a sizable inter-site interaction $V$ \cite{Seo00,Takahashi06,Seo06}.

\begin{center}
\begin{figure}
\includegraphics[width=0.9\columnwidth]{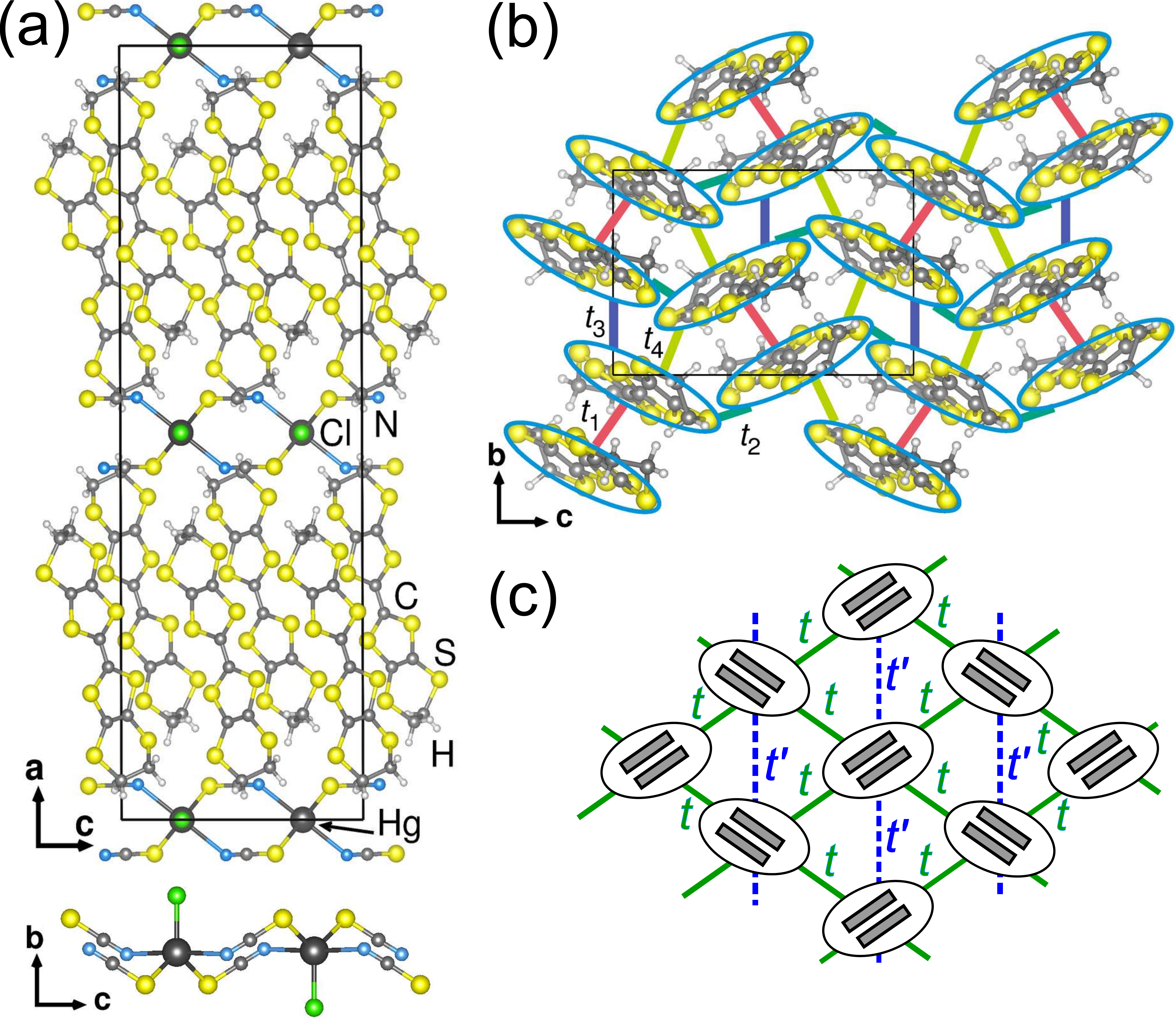}
\caption{(a) Crystal structure of \mbox{$\kappa$-(BEDT-TTF)$_2$Hg(SCN)$_2$Cl} along the out-of-plane $a$ axis (top) and side view on the anion layer (bottom) \cite{Drichko14}; (b) View on the (ET)$_2^+$ plane showing the typical $\kappa$-type arrangement of molecules. The cyan ellipses surround single ET molecules. Two parallel-aligned ET molecules form dimers. The four dominant hopping terms are denoted by $t_1$ (pink), $t_2$ (dark green), $t_3$ (blue) and $t_4$ (light green); (c) Schematic of the \textit{effective-dimer} model with hopping parameters $t$ (green) and $t^\prime$ (blue).}
\label{fig:structure}
\end{figure}
\end{center}

More recently, the research in this area has gained a new twist by the observation of strong hints for ferroelectricity in some dimerized ET-based materials \cite{Yamamoto08, AbdelJawad10, Iguchi13, Lunkenheimer15}. This came as a surprise as these systems have been primarily discussed in the so-called \textit{dimer-Mott} limit \cite{Kino95,Kanoda97,Powell11}, where the Mott insulating state is solely driven by a strong $U$, and lacks a CO instability. In this limit, (ET)$_2$ dimers are considered as single sites due to a strong intermolecular interaction $t_1$ (cf.\,Fig.\,\ref{fig:structure}(b)), being much larger than the inter-dimer interactions $t$ and $t'$ (Fig.\,\ref{fig:structure}(c)). This results in a 1/2-filled band, in which intradimer charge degrees of freedom are completely frozen. However, remarkably, for \mbox{$\kappa$-(ET)$_2$Cu[N(CN)$_2$]Cl}, ferroelectric order was found at $T_{FE}$ \cite{Lunkenheimer12, Lang14, Lunkenheimer15b} which coincides with long-range antiferromagnetic (afm) order \cite{Miyagawa95} at $T_N\,\simeq\,T_{FE}$. It has been suggested that in these dimerized systems the electric dipoles originate from CO \cite{AbdelJawad10, Lunkenheimer12, Hotta10, Naka10, Clay10, Goki13, Kaneko17}, i.e., a charge disproportionation by $\pm \delta$ within the ET dimers, suggesting an essential breakdown of the dimer-Mott scenario. However, this view has been challenged as a definite proof of CO for this family of dimerized ET systems is still missing\cite{Sedlmeier12,Tomic13,Pinteric16}.

 In this Letter, we provide evidence for an electronically-driven ferroelectricity in the related dimerized salt \mbox{$\kappa$-(ET)$_2$Hg(SCN)$_2$Cl}, where CO was clearly identified by vibrational spectroscopy\cite{Drichko14,Hassan17}. Based on our density functional theory calculations, this material has a moderate strength of dimerization thus bridging the gap between 1/4-filled CO and 1/2-filled dimer-Mott systems. We demonstrate that the transition from a metal to a CO insulator in this compound at $T_{MI}\,=\,T_{CO}\,\,\approx 25-30\,$K is accompanied by the formation of ferroelectric order of order-disorder-type, where disordered electric dipoles exist already in the paraelectric phase, and become ordered below $T_{FE} = T_{MI}$. Our results highlight the role of \textit{intra}-dimer degrees of freedom in creating novel states, such as electronically-driven multiferroicity or CO-induced quasi-1D spin-liquids. In addition, our findings underline the model character of the $\kappa$-(ET)$_2X$ systems in studying the interplay of charge-, spin- and lattice\cite{Gati16}-degrees of freedom in the presence of geometrical frustration\cite{Powell11} close to the Mott transition.

\textit{Structure and ab initio-derived hopping integrals. ---} \mbox{$\kappa$-(ET)$_2$Hg(SCN)$_2$Cl}, crystallizing in the monoclinic structure\cite{Konovalikhin92,Drichko14} $C$2/$c$, consists of alternating thick layers of organic ET molecules, separated by thin anion sheets, cf.\,Fig.\,\ref{fig:structure}(a). \textit{Ab initio} density functional theory calculations were performed using the full potential local orbital (FPLO)\cite{Koepernik99} basis and generalized gradient approximation\cite{Perdew96} for the experimentally determined structure\cite{Drichko14} at room temperature. The tight binding parameters $(t_1, t_2, t_3, t_4)$ (see Fig.\,\ref{fig:structure}(b)) were extracted from fits to the bandstructure. We find values at 296\,K of $t_1$ = 126.6\,meV, $t_2$ = 60.0\,meV, $t_3$ = 80.8\,meV, and $t_4$ = 42.0\,meV (see SI for $T$ dependence of the $t_i$'s). We use the usual geometric formulas $t = (t_2+t_4)/2$, $t^\prime = t_3 / 2$ for assessing the hopping parameters $t\,=\,51.0$\,meV and $t^\prime\,=\,40.4$\,meV of the effective-dimer model (see Fig.\,\ref{fig:structure} (c)).

\textit{Experiments. ---} Single crystals of \mbox{$\kappa$-(ET)$_2$Hg(SCN)$_2$Cl} were grown by electrocrystallization (see SI). Overall 4 crystals (3 for dielectric measurements, 1 for thermal expansion measurements) of two different sources (labeled with either AF or JAS) were studied to check for sample-to-sample variations. Dielectric measurements were performed with the electric field applied along the out-of-plane $a$ axis, the only possible configuration because of the distinctly lower conductance along this axis. In the low-frequency range ($\nu < 1 \, \mathrm{MHz}$), the dielectric constant $\epsilon'$ (real part of the permittivity) and the real part of the conductivity $\sigma'$ were determined using a frequency-response analyzer (Novocontrol alpha-Analyzer) and an autobalance bridge (Agilent 4980). The system's high conductivity and the small sample size cause some uncertainties in the absolute values of $\epsilon'$. Measurements of relative length change $\Delta L_i(T)/L_i\,$, with $i\,=\,a,b,c$, were performed using a home-built capacitive dilatometer\cite{Pott83} with a resolution $\Delta L_i/L_i\,\geq\,10^{-10}$.

\begin{center}
\begin{figure}
\includegraphics[width=0.9\columnwidth]{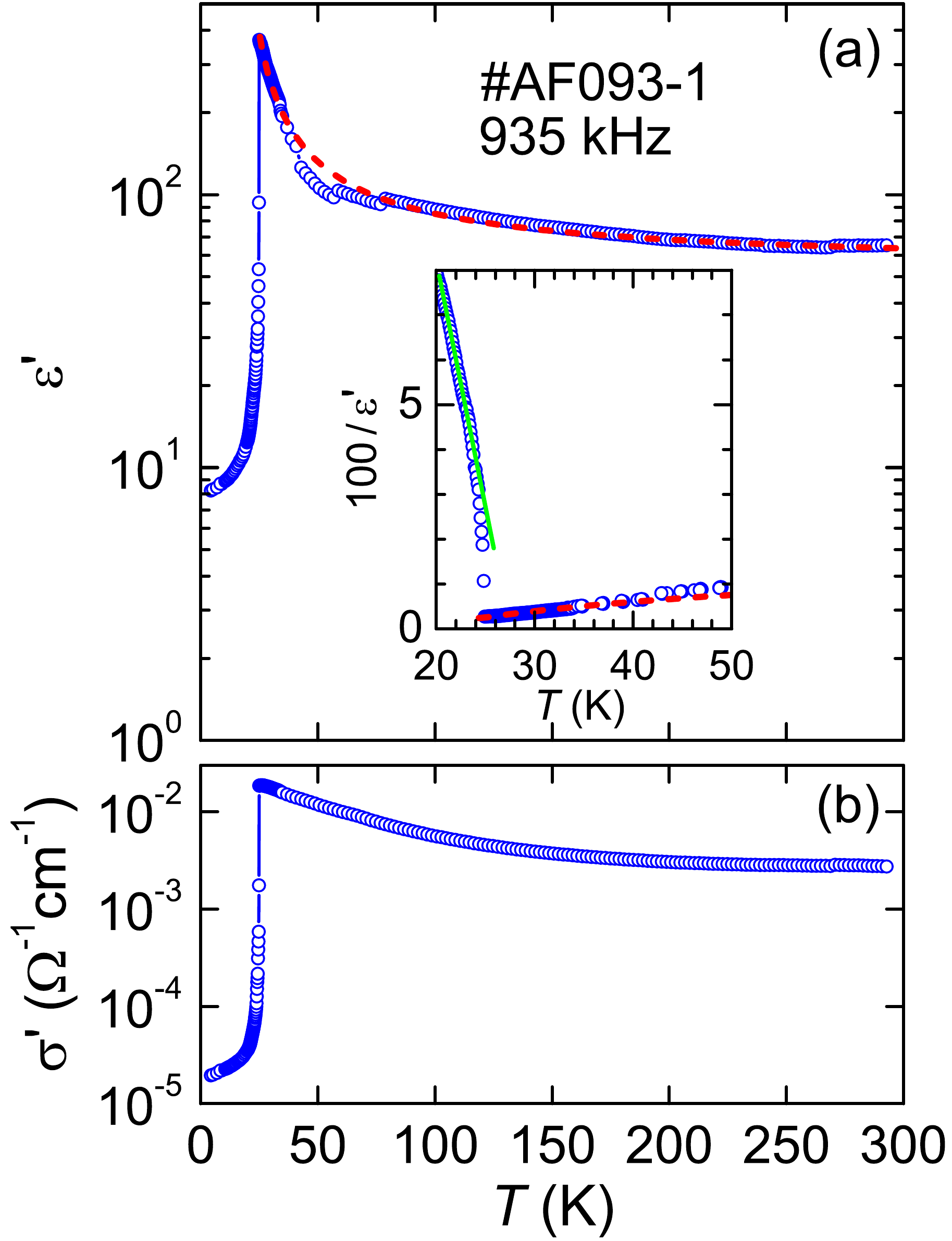}
\caption{Temperature dependence of the dielectric constant $\varepsilon'(T)$ (a) and conductivity $\sigma'(T)$ (b) of $\kappa$-(BEDT-TTF)$_2$Hg(SCN)$_2$Cl crystal \#AF093-1 measured at $935 \, \mathrm{kHz}$. Data were taken upon warming. The dashed line in (a) is a fit with a Curie-Weiss law ($T_\mathrm{CW}=17.4\,\mathrm{K}$, $C = 2500\,\mathrm{K}$) with an additional offset. The solid line connects the data points. The inset shows the inverse dielectric constant; the lines correspond to Curie-Weiss behavior.}
\label{fig:dielectric}
\end{figure}
\end{center}

Figure \ref{fig:dielectric} shows the dielectric constant $\epsilon'(T)$ (a) and the real part of the conductivity $\sigma'(T)$ (b) of crystal \#AF093-1. We find an increasing $\epsilon'(T)$ with decreasing temperature, culminating in a sharp peak at $T_{FE} \approx 25\,\mathrm{K}$, indicative of a ferroelectric transition (peak value $\approx 400$). As shown by the dashed line in Fig.\,\ref{fig:dielectric}(a), this increase can be well described by a Curie-Weiss law, $\epsilon'-\epsilon_{off}=C/(T-T_\mathrm{CW})$, with a Curie-Weiss temperature $T_\mathrm{CW}=(17\,\pm\,2)\,\mathrm{K}$ and an offset $\epsilon_{off}$, likely of extrinsic nature. The relatively small magnitude of the Curie constant of $C=(2500\,\pm\,600)\,\mathrm{K}$ is consistent with order-disorder ferroelectricity\cite{footnote4} while $C$ for displacive ferroelectrics\cite{Lines77} is usually of the order of $10^5$. By using a simple expression\cite{footnote3} to relate the Curie constant to the size of an individual dipole\cite{Mason48} $p$, we find $p\,\approx\,0.4 e d$, with $e$ the electronic charge and $d\,\approx\,4.0\,$\AA\,the distance between two ET molecules within the dimer. In light of the strong simplifications involved in this relation and the experimental uncertainties associated with the absolute values of $\epsilon'$, this value of $p$ is in reasonable agreement with the expected out-of-plane dipole moment of $0.13 e d$ created by the observed charge disproportionation\cite{Drichko14} of $\pm$ 0.1$e$ and the relative shift of the molecules within the dimer resulting in a tilt of the dipole moment by $\approx\,$50$^\circ$ with respect to the out-of-plane $a$ axis. Below $25\,\mathrm{K}$, $\epsilon'(T)$ exhibits an abrupt drop and levels off at $\epsilon' \approx 8$ at low temperatures. By looking at the inverse dielectric constant in the inset of Fig.\,\ref{fig:dielectric}, we find that, in a limited temperature range, $\epsilon'(T)$ for $T\,<\,T_\mathrm{FE}$ can also be described by a Curie-Weiss behavior (solid line), albeit with a distinctly larger slope $|d(1/\epsilon)/dT|$.

Corresponding measurements on a second crystal (\#JAS1721) from a different source, performed with a different measurement device, revealed qualitatively similar behavior with $T_\mathrm{FE} \approx 30\,\mathrm{K}$ (see SI, Fig.\,3). We did not observe any significant frequency dependence of the dielectric properties for frequencies below about 1\,MHz (see SI, Fig.\,2). However, in high-frequency measurements up to about 1\,GHz (see SI, Fig.\,4), we found an increasing suppression of the peak in $\epsilon'(T)$ with increasing frequency that resembles the typical behavior of order-disorder ferroelectrics \cite{Lines77}.

The real part of the conductivity shown in Fig.\,\ref{fig:dielectric}(b) shows metallic behavior at higher temperatures. Below 25\,K, $\sigma'(T)$ rapidly drops by about three orders of magnitude, indicating that $T_{MI} \simeq T_{FE}$. Similarly, for crystals \#JAS1721 and \#AF087 (see SI, Fig. 8), we find a rapid drop in $\sigma$($T$) at $T_{MI}\,\simeq\,T_{FE}\,\sim\,30$\,K. These findings, which are in good qualitative accord with literature results\cite{Yasin12}, provide additional evidence that the dielectric measurements indeed detect the intrinsic sample properties.

The characteristics of $\epsilon'(T)$, revealed in Fig.\,\ref{fig:dielectric}(a) (and SI Fig.\,3(a)), are remarkable in terms of the following aspects. First, the phenomenology closely resembles textbook examples of first-order ferroelectric transitions reported, e.g., for \mbox{BaTiO$_3$} or \mbox{AgNa(NO$_2$)$_2$} (Refs.\,\citen{Lines77,Johnson65,Gesi70}). This includes a Curie-Weiss temperature dependence both above and below $T_{FE}$ with strongly different slopes $|d(1/\epsilon)/dT|$, together with a Curie-Weiss temperature $T_\mathrm{CW} < T_{FE}$. Second, the observed temperature (Fig.\,\ref{fig:dielectric}(a) and SI Fig.\,3(a)) and frequency dependences (SI Fig.\,2 and 4) indicate that \mbox{$\kappa$-(ET)$_2$Hg(SCN)$_2$Cl} represents an order-disorder-type ferroelectric. This contrasts with relaxor-type ferroelectricity, characterized by a pronounced frequency dependence in $\epsilon'$ \cite{Lines77,Lunkenheimer15b}. In fact, a relaxor-type ferroelectricity has been observed for the related $\kappa$-(ET)$_2$Hg(SCN)$_2$Br salt \cite{Ivek17} which also stands out by its anomalous Raman response \cite{Hassan17}. In this and in other charge-transfer salts, the relaxational response was ascribed to the dynamics of CO domain-walls or solitons \cite{Ivek10, Fukuyama17}.

We stress that, a definite proof of ferroelectricity, which usually includes measurements of polarization hysteresis or so-called positive-up-negative-down measurements \cite{Lunkenheimer12,Scott00}, was not possible for the present compound due to its rather high conductivity, especially close to $T_\mathrm{FE}$. However, taking into account the observed characteristic temperature and frequency dependences in $\epsilon'$ and the fact that very similar results in $\epsilon'(T)$ were obtained for samples from different sources, by using different devices, the present data provide strong indications for ferroelectricity in \mbox{$\kappa$-(ET)$_2$Hg(SCN)$_2$Cl}.

\begin{center}
\begin{figure}
\includegraphics[width=0.9\columnwidth]{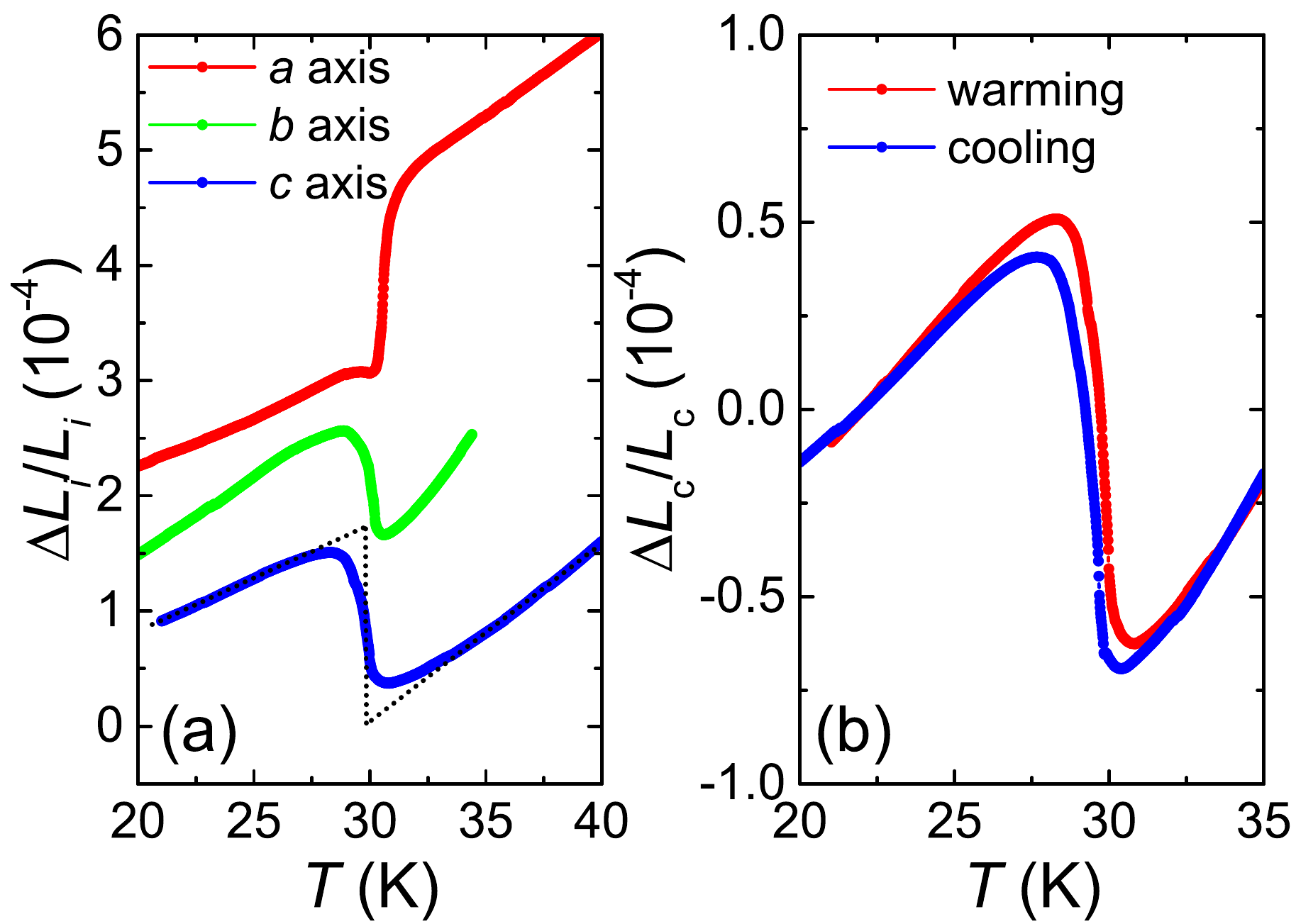}
\caption{(a) Relative length change $\Delta L_i/L_i$ vs. $T$ with $i\,=\,a,b,c$ of $\kappa$-(BEDT-TTF)$_2$Hg(SCN)$_2$Cl (crystal \#AF087-4) around the charge-order metal-insulator transition at $T_{MI}\,\approx\,30\,$K. Data were collected upon warming. The individual data sets were offset for clarity. Dotted line indicates an idealized sharp jump for the $c$-axis data. (b) Relative length change along the $c$ axis, $\Delta L_c/L_c$, around $T_{MI}$ measured upon warming and cooling.}
\label{fig:lengthchange}
\end{figure}
\end{center}

A thermodynamic investigation of the character of the CO transition is provided by measurements of the relative length change $\Delta L_i(T)/L_i$. Figure \ref{fig:lengthchange}\,(a) shows the result of $\Delta L_i(T)/L_i$ along the out-of-plane $a$ axis (see Fig.\,\ref{fig:structure}) and the two in-plane $b$ and $c$ axes around 30\,K. We observe pronounced, slightly broadened jumps in the sample length along all three axes at $T_{MI}\,\simeq T_{CO}\,\approx\,30$\,K (see SI for a detailed determination of $T_{MI}$ from the present data set). The jump-like anomalies in $\Delta L_i(T)/L_i$ and the observation of thermal hysteresis between warming and cooling (Fig.\,\ref{fig:lengthchange}\,(b)) are clear signatures of the first-order character of the CO transition, consistent with the conclusion drawn above from the $\epsilon'(T)$ results.

\begin{center}
\begin{figure}
\includegraphics[width=\columnwidth]{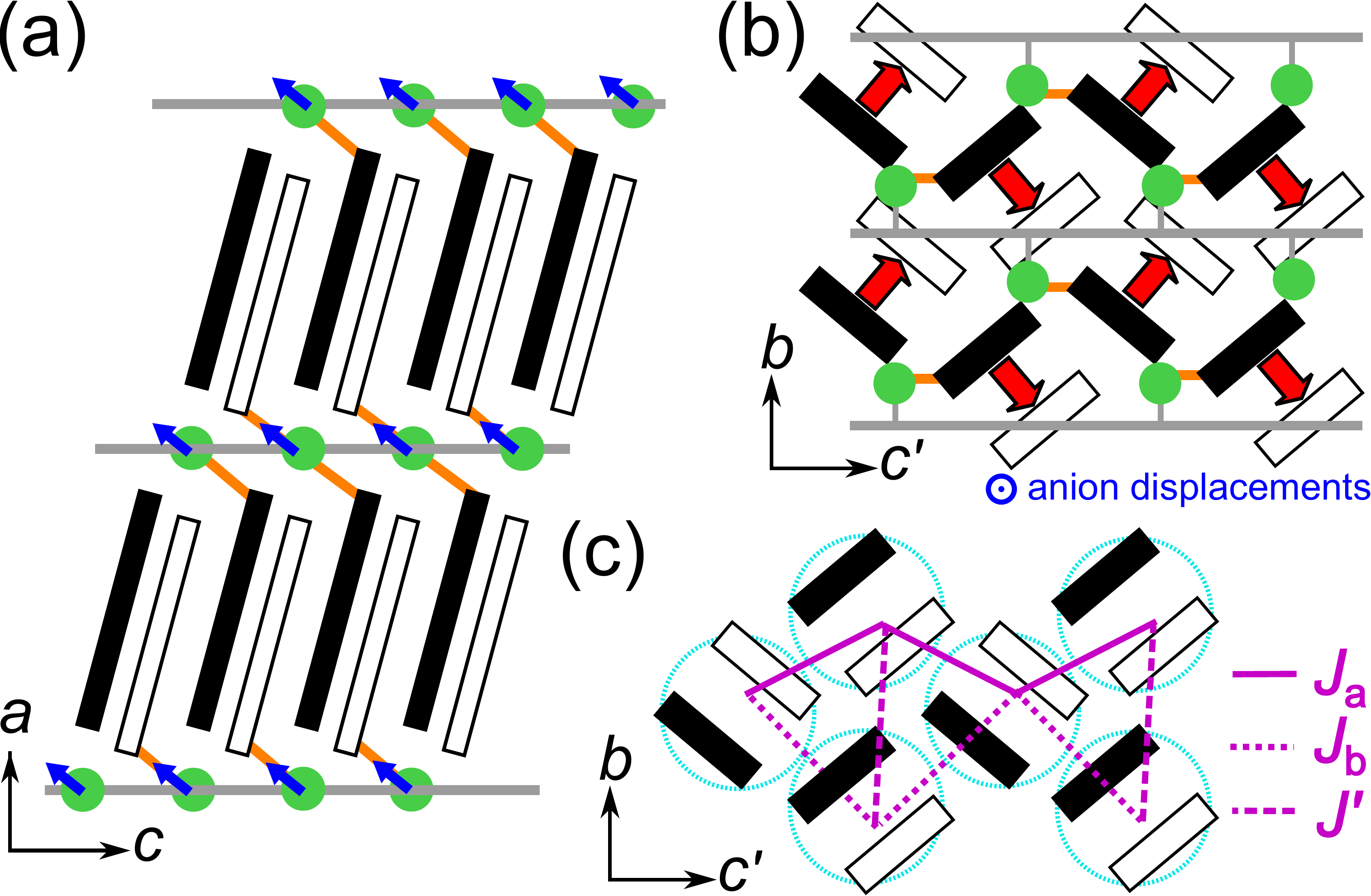}
\caption{Proposed CO pattern in $\kappa$-(BEDT-TTF)$_2$Hg(SCN)$_2$Cl, viewed within the $ac$ plane (a) and the $bc^\prime$ plane (b). $c^\prime$ accounts for a small rotation of the $c$ axis with respect to the $a$ axis due to the inclination of the BEDT-TTF molecules. Grey lines correspond to the anions, green circles correspond to Hg(SCN)$_2$Cl units of the anion layer. White (black) rectangles correspond to charge-rich (charge-poor) BEDT-TTF molecules with charge $0.5+\delta$\,($0.5-\delta$) in the charge-ordered state. Orange solid lines illustrate the interaction path between the S and H atoms in the BEDT-TTF layer and the Cl and C atoms in the anion layer. Blue arrows indicate the shift of the anions in response to the charge order in the BEDT-TTF layer. Thick red arrows indicate electric dipoles $p$. (c) Dominant magnetic exchange-coupling paths $J_a$, $J_b$ and $J^\prime$ (see main text) for the proposed CO state.}
\label{fig:chargeorderpattern}
\end{figure}
\end{center}

Surprisingly, the dominant lattice response to CO is found along the out-of-plane $a$ axis, yielding a pronounced decrease upon cooling below $T_{CO}$. This highlights a strong involvement of the anion layer in the formation of the CO state as a result of the ionic character of the material: The change in the charge distribution within the ET layers from a homogeneous distribution above $T_{CO}$ to a charge-modulated state below will necessarily induce finite shifts of the counterions in the anion layer\cite{footnote2}. We therefore include the anions in the discussion of possible CO patterns in \mbox{$\kappa$-(ET)$_2$Hg(SCN)$_2$Cl} in analogy to Ref.\,\citen{deSouza08}. Figure \ref{fig:chargeorderpattern} shows a schematic view of the structure of \mbox{$\kappa$-(ET)$_2$Hg(SCN)$_2$Cl} including the ET molecules (rectangles) and the nearby anion layers. The anions form a chain-like structure along the $c$ axis (grey lines in Fig.\,\ref{fig:chargeorderpattern}) with short-side chains formed by the terminal ligand Cl (green circles) along the $b$ axis. For the following discussion, we assume that the charge order modifies the electrostatic interactions between the cations and anions, which involve close contacts between the electropositive S and H atoms in the donors, and the electronegative Cl and C atoms in the anion layer. Through these interactions, each (Hg(SCN)$_2$Cl)$^-$ unit in the anion chain is linked to two ET molecules belonging to different layers (shown schematically by orange lines in Fig.\,\ref{fig:chargeorderpattern}(a)). Above $T_{CO}$ the charge is homogeneously distributed on the ET molecules ($\delta$ = 0), corresponding to a charge of $+0.5\,e$ per ET. Thus, the position of the Cl$^{-}$ ion is symmetric with respect to the surrounding (ET)$^{+0.5}$ molecules. Upon cooling through $T_{CO}$, the charge distribution is modulated by $\pm\,\delta$ within the ET layer\cite{Drichko14}. As a consequence, in order to minimize the overall Coulomb energy, the anions slightly shift towards the charge-rich sites (white rectangles), as indicated by the blue arrows in Fig.\,\ref{fig:chargeorderpattern}(a). As this motion, which results in a dominant effect along the $a$ axis, is uniform for all chains it identifies the CO pattern unambiguously, cf.\,Fig.\,\ref{fig:chargeorderpattern}(a) and (b). In the resulting CO pattern the charge-rich molecules are arranged in stripes along the $c$ axis and alternate with charge-poor stripes along the $b$ axis (see Fig.\,\ref{fig:chargeorderpattern}(b)). This CO pattern is consistent with the suggestion put forward in Ref.\,\citen{Drichko14} based on the anisotropy of conductivity spectra. We stress that this type of CO breaks the inversion symmetry both within and between the layers, ensuring long-range 3D ferroelectric order. However, the conclusions on the structural change at the CO transition are only speculative at present, as X-ray investigations at 10\,K\cite{Drichko14}, aimed at detecting the CO pattern, failed to resolve the predicted symmetry-breaking shifts.

For discussing these results in the wider context of dimerized (ET)$_2$X materials, we use the ratio $t_1/t'$ to quantify the strength of dimerization. In the limit of weak or no dimerization, a non-magnetic CO ground state is adopted, as has been well established in $\theta$-phase salts\cite{Kuroki09}. On the other hand, for \mbox{$\kappa$-(ET)$_2$Cu[N(CN)$_2$]Cl}, where $t_1/t' \sim$ 5 reflects a relatively strong dimerization\cite{Kandpal09,Guterding15,Guterding16a}, the notion of a dimer-Mott insulating state\cite{Kanoda97,Kino95,Powell11} has been widely used, and the existence of CO as the origin of the observed ferroelectricity has been debated\cite{Sedlmeier12,Tomic13,Pinteric16}. Hence, the present \mbox{$\kappa$-(ET)$_2$Hg(SCN)$_2$Cl} system, with $t_1/t' \sim$ 3, being located in the middle between these two extreme cases, may provide the key for a better understanding of the physics in the wide class of dimerized (ET)$_2$X materials. Our finding of ferroelectricity in \mbox{$\kappa$-(ET)$_2$Hg(SCN)$_2$Cl}, which is most likely driven by the observed CO within the ET dimers \cite{Drichko14}, clearly demonstrates the importance of intra-dimer charge degrees of freedom in these materials. Hence, the minimal model able to capture theses effects has to include two molecular orbitals on each dimer and a $3/4$ band filling. In fact, by using the hopping parameters $t_1$...$t_4$ relevant for the rather strongly dimerized \mbox{$\kappa$-(ET)$_2$Cu[N(CN)$_2$]Cl}, and by using an extended two-orbital Hubbard model on a triangular lattice at 3/4-electron filling, Kaneko \textit{et al.}\cite{Kaneko17} recently revealed the possibility for a CO ground state for this material, pointing to the relevance of intra-dimer degrees of freedom even for stronger dimerization.

In light of the peculiar multiferroic state with $T_{FE} \sim T_N$ proposed for $\kappa$-(ET)$_2$Cu[N(CN)$_2$]Cl, one may ask how charge order interacts with the magnetic degrees of freedom in the present \mbox{$\kappa$-(ET)$_2$Hg(SCN)$_2$Cl} material. Initially, Yasin \textit{et al.} \cite{Yasin12} suggested afm order to coincide with $T_{CO}$ in \mbox{$\kappa$-(ET)$_2$Hg(SCN)$_2$Cl}, based on the result of electron spin resonance (ESR) measurements. However, as discussed in detail in the SI, our own ESR investigations along with specific heat measurements fail to reveal any clear signature of a magnetic transition around $T_{MI}$. Naively, one may assign the absence of long-range magnetic order to the geometric frustration, inherent to the $\kappa$-type triangular arrangement of dimers. In fact, for the frustration paramter $t^\prime/t$ we find 0.79 in the effective-dimer model, which neglects CO, - a value significantly larger than $t^\prime/t\,\sim\,$0.43 for \mbox{$\kappa$-(ET)$_2$Cu[N(CN)$_2$]Cl}. However, charge order must have an effect on the local magnetic interactions due to the redistribution of charge within each dimer \cite{Naka10,Naka16}, i.e., in first approximation $J_i\,\propto\,t_i^2(1\,\pm\,A_i\delta)$ with a proportionality constant $A_i$. Following Naka and Ishihara \cite{Naka10,Naka16}, we anticipate that the CO pattern in Fig.\,\ref{fig:chargeorderpattern} would enhance interactions $J_a$, while suppressing the coupling $J_b$ (see Fig.\,\ref{fig:chargeorderpattern}(c)). At the same time, $J^\prime$ would not be strongly affected by CO. We propose that this modification of the interactions may lead to an effective dimensional reduction \cite{Starykh07} due to the underlying frustration ($J^\prime \sim J_b$) which promotes a quasi-1D spin-liquid state. This novel CO-driven effect could then explain the absence of magnetic order in the present material. A crucial test of this proposal would be to probe the dimensionality of spin correlations below $T_{CO}$, via e.g. polarized Raman scattering\cite{Lemmens03, Hassan17} or thermal transport anisotropy \cite{Hess07}.

\textit{Summary. ---} Clear evidence is provided for an order-disorder type ferroelectric state in dimerized \mbox{$\kappa$-(ET)$_2$Hg(SCN)$_2$Cl}, driven by charge order within the (ET)$_2$ dimers and stabilized by a coupling to the anions. According to our \textit{ab initio} density functional theory calculations, this material is characterized by a moderate strength of dimerization $t_1/t^\prime \sim$ 3. Our results highlight the role of intra-dimer degrees of freedom in dimerized (ET)$_2X$ materials in promoting intriguing states. Besides the possibility for electronically-driven multiferroicity, we propose for the present material that charge order in the presence of strong frustration may induce a quasi-1D spin-liquid state as a consequence of dimensional reduction.

\begin{acknowledgments}
This work was supported by the Deutsche Forschungsgemeinschaft through the Transregional Collaborative Research Centers TR49 and TRR 80. JAS acknowledges support from the Independent Research and Development program from the NSF while working at the Foundation. We thank Ryui Kaneko for theoretical input, and Mamoun Hemmida, Martin Dressel and Tomislav Ivek for useful discussions of the magnetic properties.
\end{acknowledgments}

\bibliographystyle{apsrev}
\bibliography{Lit} 

\begin{thebibliography}{62}
\bibitem[{foo({\natexlab{a}})}]{footnote4}
\bibinfo{note}{The absence of any indications for a displacive-type of
  ferroelectricity implies that the dielectric response can be assigned solely
  to the electric dipoles within the ET layers without any contribution from
  the alternating anion-cation system.}
\bibitem[{foo({\natexlab{c}})}]{footnote2}
\bibinfo{note}{{The same conclusion was drawn from the anisotropic lattice
  effects revealed at the CO transition in the quasi-1D (TMTTF)$_2X$
  salts\cite{deSouza08}.}}
\end{thebibliography}

\clearpage


\section*{Supplementary material}

# Supplementary Information: Evidence for electronically-driven ferroelectricity in a strongly correlated dimerized BEDT-TTF molecular conductor

Elena Gati, Jonas K.H. Fischer, Peter Lunkenheimer, David Zielke, Sebastian Köhler, Felizitas Kolb, Hans-Albrecht Krug von Nidda, Stephen M. Winter, Harald Schubert, John A. Schlueter, Harald O. Jeschke, Roser Valentí, and Michael Lang
(Dated: June 30, 2018)

## I. METHODS - DETAILS

*Ab initio calculations*: We calculated the bandstructures of $\kappa$-(ET)$_2$Hg(SCN)$_2$Cl for the staggered majority conformation of the ethylene endgroups and relaxed all hydrogen positions. All published structures from $T = 10$ K to $T = 296$ K have some degree of ethylene end group disorder, which has significant consequences for the electronic structure[1]. The staggered ethylene end group configuration predominates at all temperatures, with 85% staggered ET molecules at $T = 10$ K, 93% at $T = 50$ K, 84% at $T = 100$ K. At $T = 296$ K, both endgroups are significantly disordered. We also find the structures with staggered ethylene endgroup configuration to be significantly lower in energy. We relaxed all hydrogen positions as they are computed rather than measured in the experimental structures[2]. We extract a tight binding representation of the four bands at the Fermi level[3] (see Supplementary (SI) Fig. 1), considering each ET molecule as a site[4]. We use projective Wannier functions as implemented in FPLO[5].

*Single crystal growth*: In order to check for sample-to-sample variations, crystals of two different sources were used. Crystals were grown following the procedure reported in Ref. 6. Only for those crystals labeled with #AF087 (#AF093) the following minor modifications were applied: Pure TCE (1,1,2-Trichloroethane) was employed as a solvent with a mixture of Hg(SCN)$_2$ and PPNCl (bis(triphenylphosphoranylidene)ammonium chloride) in a molar ratio of 1:1 serving as the electrolyte. The electrolyte was given in a ten-fold excess to the solution in relation to the ET. A constant current of $0.2\,\mu$A ($0.3\,\mu$A) was applied to platinum electrodes, resulting in a voltage of 0.1 V - 0.3 V. Crystal growth was performed at a temperature of 20 °C and crystals were collected after 4-5 weeks. The crystals used for this work had typical dimensions of $1\,\mathrm{mm}^2 \times 0.35\,\mathrm{mm}$ (#AF087), $0.25\,\mathrm{mm}^2 \times 0.05\,\mathrm{mm}$ (#AF093) and $1\,\mathrm{mm}^2 \times 0.5\,\mathrm{mm}$ (#JAS1721).

*Dielectric measurements*: For the dielectric measurements, electrodes of graphite paste were applied on either side of the plate-like crystals. Due to the relatively small sample sizes and, hence, small absolute values of the measured capacitances, stray capacitances may lead to an additive contribution in $\varepsilon'$, especially in temperature/frequency regions where the intrinsic $\varepsilon'$ is small. As the absolute values of $\varepsilon'$ are not relevant in the context of the present work, no efforts have been made to correct for this contribution. Additional measurements at high frequencies (1 MHz $< \nu <$ 1 GHz) were performed by a coaxial reflection technique employing an impedance analyzer (Keysight E4991B)[7]. Sample cooling with rates of $\pm (0.1$ - $0.4)$ K/min was achieved by a $^4$He-bath cryostat (Cryovac).

*Measurements of relative length change*: Measurements of the relative length change $\Delta L_i(T)/L_i = (L_i(T) - L_i(T_0))/L_i(300\,\mathrm{K})$, with $L_i$ being the length along the axis $i$ and $T_0$ a reference temperature, were performed using a home-built high-resolution capacitive dilatometer. Measurements were performed upon heating and cooling using a rate of $\pm 1.5$ K/h. Prior to the measurements, the sample was cooled slowly with $-3$ K/h through the glass-transition region[8] around $T_g \approx 63$ K. Crystals were oriented by eye resulting in a maximum misalignment of 5°.

## II. BAND STRUCTURE CALCULATIONS

In Figure 1 we show the evolution of the calculated electronic band structure and density of states with temperature. Note, that 10 K and 50 K structures have been determined based on x-ray measurements at a different instrument; we consider 100 K and 296 K structures more reliable. Therefore, only limited comparison between the two sets of structural data is possible. We find that there is no qualitative change in the electronic structure and Hamiltonian parameters between the original structures and those with equilibrium hydrogen positions.

## III. ADDITIONAL DIELECTRIC-SPECTROSCOPY MEASUREMENTS

### A. Low-frequency results

Figure 2 shows the temperature dependence of $\epsilon'$ of crystal #AF093-1 for additional frequencies as measured by the frequency-response analyzer. No significant frequency dependence is detected.

TABLE I. Tight binding parameters of $\kappa$-(BEDT-TTF)$_2$Hg(SCN)$_2$Cl at four different temperatures, calculated with relaxed hydrogen positions and for the majority endgroup configurations.

| $T$ (K) | $t_1$ (meV) | $t_2$ (meV) | $t_3$ (meV) | $t_4$ (meV) | $t'/t$ | $U/t$ |
|---|---|---|---|---|---|---|
| 10 | 117.4 | 62.7 | 88.1 | 42.2 | 0.840 | 4.5 |
| 50 | 114.5 | 63.1 | 89.5 | 40.3 | 0.865 | 4.4 |
| 100 | 118.9 | 64.4 | 89.1 | 41.6 | 0.841 | 4.5 |
| 296 | 126.6 | 60.0 | 80.8 | 42.0 | 0.792 | 5.0 |

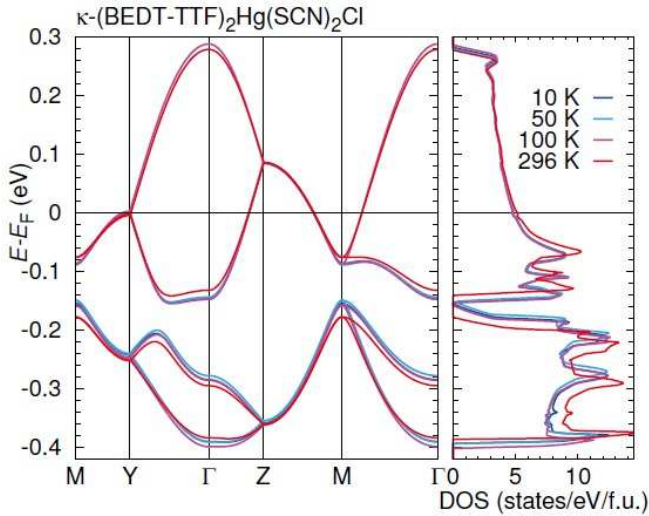

FIG. 1. Band structure of $\kappa$-(BEDT-TTF)$_2$Hg(SCN)$_2$Cl at four different temperatures. The majority ethylene endgroup configuration has been chosen in all cases. Note that the underlying structural data at $T = 10$ K and 50 K had been determined based on measurements at a different instrument; we consider the 100 K and 296 K data more reliable.

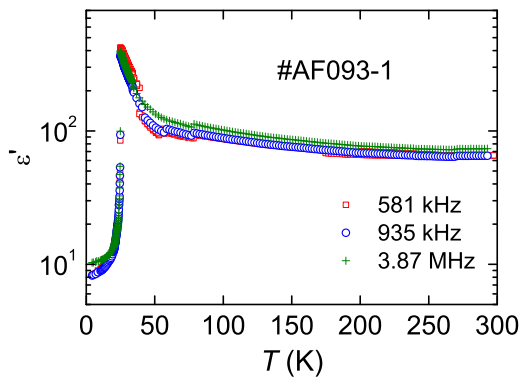

FIG. 2. Temperature dependence of the dielectric constant $\epsilon'$ of $\kappa$-(BEDT-TTF)$_2$Hg(SCN)$_2$Cl, crystal #AF093-1 measured at three different frequencies.

Figure 3 shows the dielectric constant $\varepsilon'$ (a) and conductivity $\sigma'$ (b) of crystal #JAS1721. The measurements were performed using an autobalance bridge instead of the frequency-response analyzer employed for sample #AF093-1 (Fig. 2 in the main text). Just as for the latter, a well-pronounced asymmetric peak is observed (in this case at 30 K), exhibiting the typical signature of a first-order ferroelectric transition. The peak value of 500 is somewhat higher but of similar order of magnitude as for sample #AF093-1. Its high-temperature flank again can be reasonably well fitted by a Curie-Weiss law with $C = (6300 \pm 1000)$K (dashed line in Fig. 3). In the fitting procedure, $T_{\rm CW}$ was fixed at the same value of 17 K as found for sample #AF093-1. Just as for the latter [Fig. 2(b) in the main text], the conductivity of crystal #JAS1721 [Fig. 3(b)] shows weakly temperature-dependent metal-like behavior above $T_{\rm FE} = T_{\rm MI}$ and a sharp drop by several decades below. For this sample this drop reaches almost five orders of magnitude. Apart from the sample-to-sample variation of the electrical properties, the overall behavior of crystals #AF093-1 and #JAS1721 is similar and both show similar signatures of a ferroelectric transition.

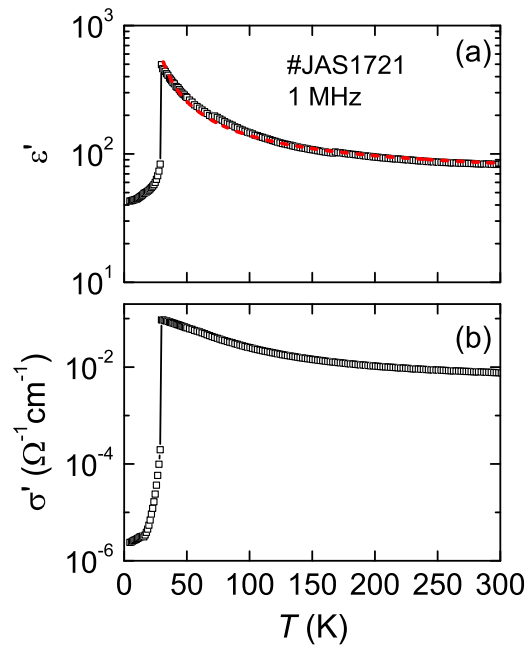

FIG. 3. Temperature dependence of the dielectric constant $\varepsilon'$ (a) and conductivity $\sigma'$ (b) of $\kappa$-(BEDT-TTF)$_2$Hg(SCN)$_2$Cl, crystal #JAS1721 measured at 1 MHz. The dashed line has the same meaning as in Fig. 2 of the main text.

## B. High-frequency results

Figure 4 shows the temperature-dependent dielectric constant of crystal #AF093-2 at high frequencies, $\nu \geq 4.35$ MHz, measured with a coaxial reflection technique using an impedance analyzer[7]. For the lowest frequencies, the typical asymmetric peak, as also detected in the low-frequency measurements of Figs. 2(a) in the main text and 3(a), is revealed (cf. $\varepsilon'(T)$ of crystal #AF093-1 shown by the open circles in Fig. 4). It becomes successively suppressed with increasing frequency. Finally, at the highest frequencies of several hundred MHz, a minimum develops. A similar phenomenon was also observed in other order-disorder ferroelectrics and can be explained by an interplay of the strongly increasing static dielectric constant and the slowing down of the dipolar dynamics when approaching the transition from above $T_{\text{FE}}$[9–11]. An alternative dispersion mechanism was proposed in a recent work[12] which theoretically predicts excitations arising from solitons in quarter-filled molecular solids, however at lower frequencies in the 10 kHz range. When closely inspecting Fig. 4, a small kink is seen at about 25 K, slightly below the peak temperature. This seems to be an artifact as it was only observed in this crystal and may indicate the existence of two sample regions with slightly different $T_{\text{FE}}$.

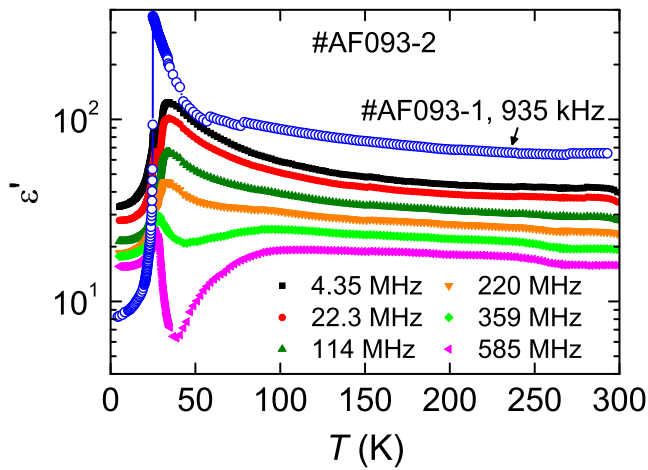

FIG. 4. Temperature dependence of the dielectric constant of $\kappa$-(BEDT-TTF)$_2$Hg(SCN)$_2$Cl, crystal #AF093-2, measured for several high frequencies up to 585 MHz (closed symbols). For comparison, the result for crystal #AF093-1 obtained at 935 kHz is shown (open circles).
.

## IV. DETERMINATION OF $T_{MI}$ FROM MEASUREMENTS OF THE RELATIVE LENGTH CHANGE $\Delta L_i/L_i$

In order to determine the transition temperature $T_{MI}$ from the data sets of the relative length change $\Delta L_i(T)/L_i$ (see Fig. 3(a) in the main text), we assigned the inflection point of each $\Delta L_i(T)/L_i$ to $T_{MI}$. This procedure yields slightly different $T_{MI}$ values for the data sets taken along the axes $i = a, b, c$ as already apparent from the bare data. We find $T_{MI,a} = (30.5 \pm 0.1)$ K, $T_{MI,b} = (30.0 \pm 0.1)$ K and $T_{MI,c} = (29.8 \pm 0.1)$ K. This small discrepancy can be assigned to the influence of the small uniaxial pressure exerted by the dilatometer when mounting the crystal. The corresponding force (typically less than 1 N) is caused by the flat springs[13] which suspend the upper plate of the capacitor in the dilatometer cell. For the present crystal of dimensions $0.25 \times 1.1 \times 0.9$ mm$^3$, the corresponding uniaxial pressure components $P_i$ amount to $P_a \approx 0.6$ MPa and $P_b \approx P_c \approx 2$ MPa. As deduced from the different signs in the expansion $\Delta L_i/L_i|_{T_{MI}}$ (see Fig. 3, main text), the uniaxial pressure dependence of $T_{MI}$ is positive for uniaxial pressure along the out-of-plane $a$ axis and negative for the other two directions. Correspondingly, as a result of the uniaxial pressure acting on the crystal along the measuring direction, we find a value for $T_{MI,a}$ which is slightly larger than those of $T_{MI,b}$ and $T_{MI,c}$. In addition, using the criterion defined above, we can quantify the hysteresis presented in Fig. 3(b) in the main text. We find a difference in the transition temperature determined from data sets taken upon warming and cooling of $\Delta T_{MI} = (0.3 \pm 0.1)$ K.

## V. ENTROPY RELEASE AT THE CHARGE-ORDER METAL-INSULATOR TRANSITION

In the following, we estimate the entropy release at the charge-order metal-insulator transition at $T_{MI}$ by using results from specific heat $C(T)$ and the relative length change $\Delta L_i(T)/L_i$.

### A. Specific heat - Experimental details

Measurements of specific heat were performed by employing a high-resolution ac-modulation technique[14] on a single crystal from batch #AF087-3 of mass $m = (75 \pm 20)\,\mu$g. Details of the setup, specially designed for measuring small plate-like crystals, are presented in Ref. 15. Measurements were performed upon warming in the range $2\,\text{K} \leq T \leq 35\,\text{K}$.

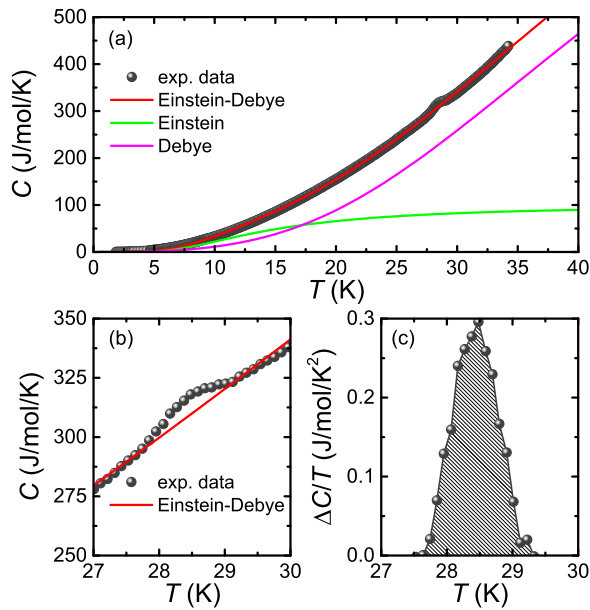

FIG. 5. Specific heat $C$ of $\kappa$-(BEDT-TTF)$_2$Hg(SCN)$_2$Cl: (a) experimental data of $C$ vs. $T$ (grey circles) in the range $2\,\mathrm{K} \leq T \leq 35\,\mathrm{K}$. Red line corresponds to a fit of the phononic background contribution $C_{Ph}$ which is described by a sum of an Einstein (green line) and a Debye contribution (pink line); (b) Anomaly, visible in the $C$ vs. $T$ data at $T \approx 28.5\,\mathrm{K}$ on expanded scales and the phononic background fit (Einstein-Debye fit); (c) Plot of $\Delta C/T = (C - C_{Ph})/T$ vs. $T$. The shaded area corresponds to the entropy change $\Delta S$ at the phase transition.

## B. Specific heat - Results

Figure 5 (a) shows data of $C$ vs. $T$ in the temperature range $2\,\mathrm{K} \leq T \leq 35\,\mathrm{K}$. We observe an overall increase of $C$ with increasing $T$ with a small hump around $T \approx 28.5\,\mathrm{K}$. We assign this feature, which is shown on an enlarged scale in Fig. 5 (b), to the signature of the metal-insulator transition at $T_{MI}$. Note that we have provided experimental evidence from measurements of the relative length change $\Delta L_i/L_i$ (see main text, Fig. 3) that this phase transition is of first-order. Consequently, a divergence in $C(T)$ at $T_{MI}$ would be expected. However, due to the ac-technique applied here, which involves temperature oscillations with typical amplitudes $\Delta T/T \sim 0.01$, sharp features can become smeared out or rounded.

In order to determine the entropy release $\Delta S$ associated with the phase transition, we have to subtract all other contributions to $C$ which evidently dominate the overall behavior of $C(T)$. As the sample is insulating for $T \leq 30\,\mathrm{K}$, electronic contributions can be neglected in this temperature regime and $C$ is assumed to be given solely by phonon contributions $C_{Ph}$. For the description of $C_{Ph}$, we employ a model in which we take Debye- and Einstein-type of oscillations into account. Whereas the former ones account well for the contribution of acoustic phonons with an, in first approximation, linear dispersion close to the Brillouin zone center, the latter one describes the contribution of optical phonons with a small dispersion. The resulting model reads as

$$C_{Ph}(T) = 9N_D R \left(\frac{T}{\Theta_D}\right)^3 \int_0^{\Theta_D/T} x^4 \frac{e^x}{(e^x-1)^2}\,\mathrm{d}x \\ + 3N_E R \left(\frac{\Theta_E}{T}\right)^2 \frac{e^{\Theta_E/T}}{(e^{\Theta_E/T}-1)^2}, \quad (1)$$

with $N_D$ ($N_E$) the number of Debye (Einstein) oscillators and $\Theta_D$ ($\Theta_E$) the Debye (Einstein) temperature. The significant contribution of optical phonons even at such low temperatures in the present case can be inferred from a plot of $C/T^3$ vs. $T$. In this representation, Debye contributions, which follow $C_{Debye} \propto T^3$ at low temperatures, give rise to a horizontal line, whereas Einstein contributions show a pronounced maximum. Such a maximum can be observed for the present data set at $T \approx 6.6\,\mathrm{K}$ reflecting the presence of low-lying optical phonons. These two dominant contributions to $C(T)$ were also identified for other charge-transfer salts, such as $\kappa$-(ET)$_2$I$_3$[16]. The low-lying optical phonons were often assigned to librational modes[16] of the anion motion[17,18].

The model contains three free parameters, namely $N_E$, $\Theta_D$ and $\Theta_E$, as we keep the total number of oscillators $N = N_D + N_E$ fixed and equate it with the number of molecules per formula unit, $N = 60$. The obtained fit, which is in very good agreement with the data set for $T < 27.5\,\mathrm{K}$, is shown in Fig. 5 (a) and (b) by the red line. The resulting fit parameters amount to $\Theta_D = (210 \pm 20)\,\mathrm{K}$, $\Theta_E = (46 \pm 5)\,\mathrm{K}$ and $N_E = 4$. The Debye temperature obtained here is in good agreement with those determined for other charge-transfer salts[16,19–21] which typically range from $180\,\mathrm{K}$ to $220\,\mathrm{K}$. Also $\Theta_E$ is similar to values found, e.g., for $\kappa$-(ET)$_2$I$_3$, the specific heat[16] of which was modeled with $N_E = 2$ oscillators. The values determined here therefore confirm the reliability of our fit.

After we determined the background, we can now proceed with analyzing the specific heat and the entropy associated with the phase transition. To this end, we evaluate $\Delta C(T)/T = (C(T) - C_{Ph}(T))/T$ (see Fig. 5 (c)). The integration of $\Delta C(T)/T$ in the temperature range from $27.5\,\mathrm{K}$ to $29.5\,\mathrm{K}$ (shaded area in Fig. 5 (c)) yields $\Delta S = (250 \pm 50)\,\mathrm{mJ\,mol^{-1}K^{-1}}$.

## C. Clausius-Clapeyron equation - Results

In the previous section, we presented the determination of the entropy change $\Delta S$ across the first-order phase transition at $T_{MI}$ from specific heat measurements. In this paragraph, we present a second, independent approach to $\Delta S$ by using the Clausius-Clapeyron equation $\mathrm{d}T_{MI}/\mathrm{d}P = \Delta V/\Delta S$ and the experimentally deter-

4

mined quantities $\Delta V$ and $dT_{MI}/dP$. To this end, we determined the volume jump[22] at the metal-insulator transition $\Delta V/V|_{T_{MI}}$ from the present $\Delta L_i/L_i$ data by $\Delta V/V|_{T_{MI}} \simeq \sum_i \Delta L_i/L_i|_{T_{MI}}$ with $i = a, b, c$. The individual $\Delta L_i/L_i|_{T_{MI}}$ contributions were determined using the following procedure (as indicated exemplarily for the $c$ axis data by the dotted lines in Fig. 3 (a), main text): The slightly broadened jumps in $\Delta L_i/L_i$ were replaced by infinitely sharp jumps by extrapolating the normal background from both sides of the transition up to $T_{MI}$. The resulting relative length changes at $T_{MI}$ amount to $\Delta L_a/L_a|_{T_{MI}} = (1.45 \pm 0.05) \cdot 10^{-4}$, $\Delta L_b/L_b|_{T_{MI}} = -(1.60 \pm 0.05) \cdot 10^{-4}$ and $\Delta L_c/L_c|_{T_{MI}} = (1.65 \pm 0.05) \cdot 10^{-4}$ and, accordingly, $\Delta V/V|_{T_{MI}} = -(1.80 \pm 0.15) \cdot 10^{-4}$. Note that a relative volume change of similar size has been revealed at the Mott metal-insulator transition for $\kappa$-(ET)$_2$Cu[N(CN)$_2$]Cl[23]. Together with the molar volume of $V_{mol} = 5.36 \cdot 10^{-4}\,\mathrm{m}^3/\mathrm{mol}$ and $dT_{MI}/dP \approx -(0.33 \pm 0.03)\,\mathrm{K/MPa}$ determined experimentally (see Fig. 8) on a crystal from the same batch, the entropy change can be determined as $\Delta S = (290 \pm 30)\,\mathrm{mJ\,mol^{-1}\,K^{-1}}$. This value of $\Delta S$ is, within the error bars, consistent with the value given in the previous section.

### D. Discussion

Given that the metal-insulator transition is a first-order transition involving only charge degrees of freedom, an entropy change of $\Delta S = \gamma T_{MI}$, with $\gamma$ being the electronic Sommerfeld coefficient, would be expected. Under this assumption, the experimentally determined $\Delta S$, gives $\gamma \simeq 10\,\mathrm{mJ\,mol^{-1}\,K^{-2}}$. This value of $\gamma$ is at the lower bound of $\gamma$ values found for other organic charge-transfer salts[16,19–21,24] which typically are in the range $10\,\mathrm{mJ\,mol^{-1}\,K^{-2}} \lesssim \gamma \lesssim 30\,\mathrm{mJ\,mol^{-1}\,K^{-2}}$. The comparably low $\gamma$ value determined here is in agreement with the somewhat weaker correlations, predicted by our DFT calculations, which results in a smaller effective mass. Thus, the present result of the entropy change $\Delta S$ can be explained consistently by considering only the effect of charge degrees of freedom. We infer that, if present at all, any change of a magnetic entropy at $T_{MI}$ would be very small.

## VI. ESR MEASUREMENTS

### A. Experimental details

Electron spin resonance (ESR) measurements have been performed in a Bruker ELEXSYS E500 spectrometer working at X− ($\nu = 9.34$ GHz) and Q-band frequencies ($\nu = 34$ GHz) with *Oxford* He-gas flow cryostats ESR 900 and ESR 935 covering the temperature range $4\,\mathrm{K} \leq T \leq 300\,\mathrm{K}$. The single crystals were fixed in high-purity *Suprasil* quartz-glass tubes by paraffin and mounted in

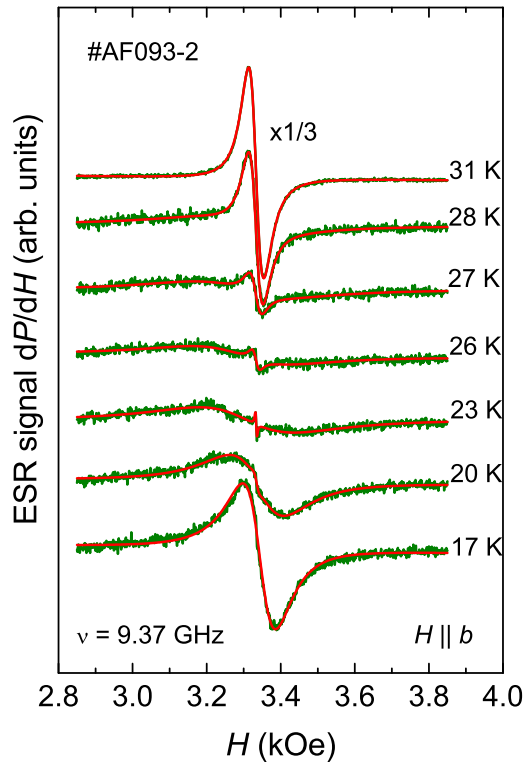

FIG. 6. Selected ESR spectra of $\kappa$-(BEDT-TTF)$_2$Hg(SCN)$_2$Cl #AF093-2 for the magnetic field applied along the crystallographic $b$ axis. The red solid lines correspond to fits with one or two Lorentz lines as discussed in the text.

the microwave cavity. A goniometer is used to adjust the orientation of the sample in the external static magnetic field $H$. ESR measures the microwave absorption from magnetic dipolar transitions excited by the transverse magnetic microwave field between the electronic Zeeman levels dependent on the static magnetic field. Resonance occurs if the microwave energy matches the Zeeman splitting, i.e., $h\nu = g\mu_B H$, where $h$ denotes the Planck constant, $\mu_B$ the Bohr magneton and $g$ is the $g$ value. Due to the lock-in technique with field modulation the field derivative $dP/dH$ of the absorbed microwave power is recorded.

### B. Results

Figure 6 shows typical ESR spectra of $\kappa$-(ET)$_2$Hg(SCN)$_2$Cl on crossing the MI transition. At temperatures above $T_{MI}$ the ESR spectrum consists of a single absorption line at a $g$ value close to the free electron value $g = 2.0023$, typical for conduction electrons. The line shape depends on the orientation of the magnetic microwave field with respect to the conducting $bc$ layers. If the magnetic microwave field

5

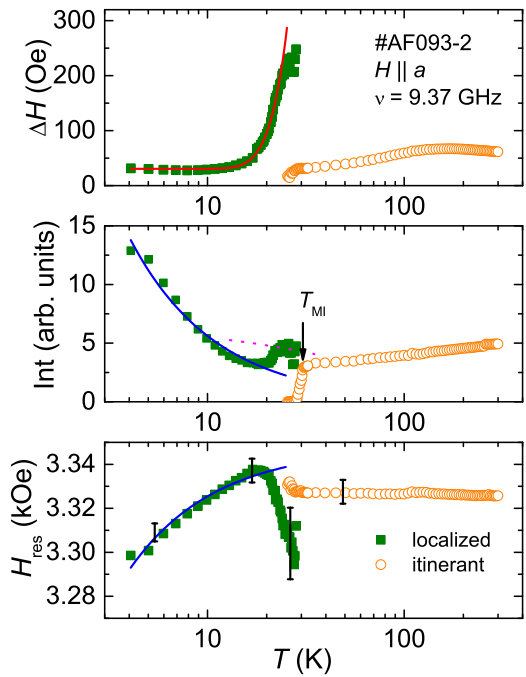

FIG. 7. Temperature dependence of linewidth $\Delta H$ (top), intensity (middle), and resonance field $H_{res}$ (bottom) of $\kappa$-(BEDT-TTF)$_2$Hg(SCN)$_2$Cl #AF093-2 for the magnetic field applied along the crystallographic $a$ axis. The red solid line in the top frame indicates a fit by an Arrhenius law with an energy gap of $\Delta/k_B \approx 130$ K, the blue solid lines in middle and bottom frame indicate pure Curie-Weiss laws while the magenta dotted line represents a Curie-Weiss (CW) law with a CW temperature of $\Theta_{CW} = -60$ K.

oscillates within the $bc$ plane one observes a pure Lorentz line, while it becomes asymmetric when the magnetic microwave field oscillates perpendicular to the $bc$ plane (not shown). In the latter case strong shielding currents are induced into the conducting planes (skin effect) leading to an admixture of dispersion into the absorption signal.[25] Hence, in general the ESR signals are fitted by the field derivative of the expression

$$P(H) = A \frac{\Delta H + \alpha(H - H_{res})}{(H - H_{res})^2 + \Delta H^2} \quad (2)$$

with the amplitude $A$, resonance field $H_{res}$, line width $\Delta H$ and dispersion-to-absorption ratio $\alpha$. For high conductivity $\alpha \to 1$ is approached, while for low conductivity ($\alpha \to 0$) the signal corresponds to a pure Lorentz line. This is the case, if the magnetic microwave field is applied within the $bc$-plane: then the shielding currents are cut along the $a$-direction, thus the skin effect can be neglected. Below $T_{MI}$ a second much broader line shows up. While the first line strongly weakens on further decreasing temperature, the second one strengthens and narrows and finally becomes dominant below 20 K.

Figure 7 exemplarily illustrates the temperature dependence of linewidth, intensity, and resonance field of both signals for the magnetic field applied along the crystallographic $a$ axis. The linewidth of the first line starts at a value of about 70 Oe at room temperature, narrows below 150 K on approaching the MI transition, below which it decreases further. Concomitantly, the second line shows up with a much broader linewidth of about 250 Oe, but strongly narrows below 20 K following an Arrhenius law with an energy gap $\Delta/k_B \approx 130$ K, and finally reaches a value of about 30 Oe at low temperatures. The (with respect to the field derivative $dP/dH$) double-integrated intensity $I \propto A\Delta H^2$ monotonously decreases with decreasing temperature down to the MI transition, below which it abruptly decreases, while the second line clearly gains intensity, decreases abruptly at 20 K and finally increases down to lowest temperature. Note that the regime between 20 and 30 K can be approximated by a Curie-Weiss law with CW temperature of about -60 K, while below 15 K a pure Curie law best approximates the data. The resonance field is temperature independent at $g_a = 2.013$ above $T_{MI}$. To lower temperatures the second broad line appears at $g \approx 2.03$ shifts slightly below 2.01 at 20 K and returns on further cooling approximately following a Curie law like the intensity. At 4 K one finds $g_a(T = 4 \text{ K}) \approx 2.03$.

For $H||b$ and $H||c$ (not shown) the temperature dependence of linewidth and intensity is very similar to that for $H||a$. The high-temperature $g$ values are found at $g_b = 2.009$ and $g_c = 2.006$. Regarding low temperatures, the resonance field decreases on approaching 4 K for $H||b$ ($g_b(T = 4 \text{ K}) = 2.04$) but increases for $H||c$ ($g_c(T = 4 \text{ K}) = 1.95$). Comparative ESR measurements at Q-band frequency did not reveal any significant differences to the X-band results and will not be discussed further here.

Our ESR results are in general agreement with the data reported by Yasin et al.[26] The temperature dependence of linewidth, intensity, and resonance fields is comparable to our findings and reveals anomalies at $T_{MI}$ and around 20 K as well. At high temperatures the anisotropy of the $g$ tensor $g_a > g_b > g_c$ is the same, only the absolute values are slightly higher (by $\Delta g \approx 0.003$) in our case. Also the evolution of the anisotropy down to $T = 4$ K is comparable to the earlier work. However, our detailed measurements employing small temperature steps around the MI transition revealed the separation of the ESR signal into two different lines below $T_{MI}$ which has not been reported before. This clearly indicates the onset of localization of the conduction electrons at $T_{MI}$: the narrow line coming from high temperatures belongs to the itinerant electrons, while the broad line appearing below 30 K arises from the localized electrons.

The ESR intensity of the narrow line increases with increasing temperature as characteristic for the spin susceptibility of an antiferromagnetically coupled two-dimensional metallic electron system, like e.g. in iron pnictides[27]. The intensity of the contribution associated

to the localized electrons reveals a weak increase upon cooling down to 20 K. This behavior would be compatible with a Curie-Weiss law between $T_{MI}$ and 20 K. Our calibration measurements prove that roughly one electron spin per formula unit contributes to the susceptibility in this temperature range. Hence the spin system behaves like a paramagnet of antiferromagnetically exchange-coupled localized spins. The $g$ value exhibits a significant shift from the conduction-electron value at $T_{MI}$ indicating the change of local fields during the localization process. At the same time the linewidth is rather broad probably due to strong charge and spin fluctuations. Below 25 K the exponential decrease of the linewidth indicates the stabilization of charge order in agreement with the steep decrease of the dielectric constant. We note that the determined gap size $\Delta/k_B = 130$ K is consistent with the charge gap determined by optical measurements[2]. A similar linewidth behavior was observed in $\beta$-Na$_{1/3}$V$_2$O$_5$ below the metal-to-insulator transition in the charge ordered phase, where the temperature dependence of the linewidth could be clearly related to the conductivity[28]. Finally, the reason for the Curie-like behavior of intensity and resonance shift below 15 K has to be addressed: Here only approximately 20% of the spins seem to contribute. Regarding the anisotropy at 4 K, Yasin et al.[26] proposed some kind of antiferromagnetic order like in the related copper compound $\kappa$-(ET)$_2$Cu[N(CN)$_2$]Cl. However, the anisotropy is by far smaller in the present compound and proportional to the spin susceptibility, i.e., the resonance shift originates from the internal field due to the magnetization of the sample. There are similarities to the behavior in $\kappa$-(ET)$_2$Hg(SCN)$_2$Br, which exhibits weakly ferromagnetic spin-glass type properties at low temperatures due to frustration and disorder.[29] Note, however that there is no charge order in the Br compound and the anisotropy is only similar in the absolute value, but its orientation with respect to the crystallographic axes is different. Hence, with the present data a final conclusion about the magnetic ground state is not possible, but long-range antiferromagnetic order seems to be very unlikely.

## VII. RESISTIVITY UNDER PRESSURE

### A. Experimental details

The electrical resistivity $\rho$ was measured in a standard four-terminal configuration using a DC current $I = 1\,\mu$A. The current was applied within the $bc$ plane using a HP3245A universal source. The voltage was read out by a Keithley 2182A nanovoltmeter. Carbon paste (Jeol Datum Ltd., Japan) was used to attach contacts to the crystals. Measurements under finite pressure were performed up to pressures of 300 MPa. To this end, the sample was placed in a CuBe pressure cell (Institute of High-Pressure Physics, Polish Academy of Sciences, Unipress Equipment Division) that was

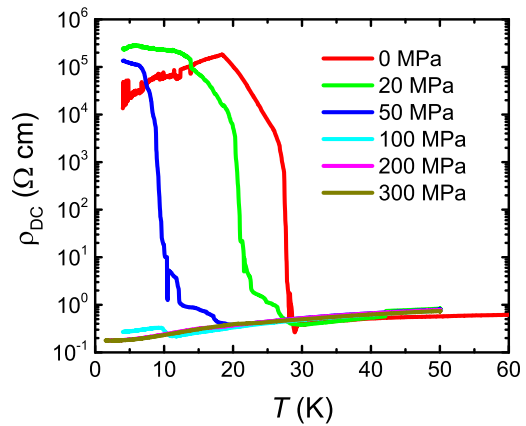

FIG. 8. Resistivity $\rho$ of $\kappa$-(BEDT-TTF)$_2$Hg(SCN)$_2$Cl (crystal #AF087-2) as a function of temperature $T$ under various external pressures ($0\,\text{MPa} \leq P \leq 300\,\text{MPa}$). Pressure values given in the legend were measured at high temperatures, i.e., 60 K. Note that the pressure medium $^4$He solidifies under $P$ at low $T$ which is accompanied by a pressure loss. This results in small jumps in the data sets taken at 100 MPa, 200 MPa and 300 MPa.

connected to a He-gas compressor via a capillary. The large volume of the compressor which is kept at room temperature ensures $P \simeq$ const. conditions during a $T$-sweep with $\Delta P = \pm 1$ MPa. Helium gas was used as a pressure-transmitting medium to ensure a hydrostatic pressure environment to lowest temperatures which is of particular importance in the study of organic charge-transfer salts. These systems usually exhibit strong and anisotropic uniaxial pressure coefficients[30] which make them sensitive to non-hydrostatic pressure components. Pressure was determined $in$-$situ$ by an $n$-InSb sensor[31]. Measurements were performed upon warming using a rate of $+7.2$ K/h.

### B. Results

First, we focus on the investigation of the properties of the charge-order MI transition. Figure 8 shows the resistivity $\rho(T)$ of a crystal of $\kappa$-(ET)$_2$Hg(SCN)$_2$Cl at ambient pressure as well as finite pressures up to 300 MPa. We note that pressure studies on this compound were reported in literature[32]. However, this study is the first one that is conducted under truly hydrostatic pressure conditions - an aspect which is of high importance whenever a phase transition is accompanied by strong and anisotropic length changes. The latter is the case for the present compound, as we will demonstrate below in more detail.

At ambient pressure the MI transition manifests itself in a sharp increase in $\rho(T)$ by 3 to 4 orders of

magnitude upon cooling below at $T_{MI} \approx 29$ K, followed by a shoulder and a less rapid increase at slightly lower temperatures, as similarly reported in Refs. 26, 32. With increasing pressure, the transition temperature becomes rapidly suppressed to lower temperatures while a broad tail, composed of several small steps, develops at the high-temperature end of the transition. Since in this temperature-pressure range the pressure-transmitting medium helium is still in its liquid phase, we can rule out that this broadening is due to non-hydrostatic pressure conditions. By assigning the transition temperature $T_{MI}$ to the point where $\mathrm{d}\log\rho(T)/\mathrm{d}T$ becomes maximum, we find an initial rate of $\mathrm{d}T_{MI}/\mathrm{d}P \approx -(0.33 \pm 0.03)$ K/MPa. This value is consistent with the pressure dependence of $T_{MI}$ revealed from resistivity measurements by using oil as a pressure-transmitting medium, although there the broadening of the transition at finite pressure was distinctly more pronounced[32].

At a pressure of 100 MPa, the data do not reveal any signature of a MI transition and the system remains metallic down to the lowest temperature of our experiment of $T = 2$ K. Note that the small step-like increase of the resistivity below about 10 K is due to the solidification of helium which is accompanied by a pressure loss from 100 MPa to about 65 MPa. We find that the metallic state which evolves at low temperature does not change much upon further increasing the pressure. Especially, there are no indications for superconductivity up to 300 MPa for $T > 2$ K. Such an insensitivity of the metallic state to pressure changes and the lack of pressure-induced superconductivity are in marked contrast to those $\kappa$-$(ET)_2 X$ salts which are located close to the Mott transition $P$[33–36].



\end{document}